\documentclass[twocolumn]{article}
\usepackage{graphicx} 
\usepackage{amsmath} 
\usepackage{amsfonts}
\usepackage{quantikz}
\usepackage{float}
\usepackage{pgfplots}
\usepackage{pgfplotstable} 
\usepackage[sorting=none,backend=bibtex,style=numeric]{biblatex}
\usepackage{hyperref}

\usepackage{caption}
\usepackage{subcaption}

\pgfplotsset{compat=1.18}

\newcommand{\imagePath}{images/}

\newcommand{\PCBO}{CPBO}
\newcommand{\PUBO}{PBO}
\newcommand{\pa}[1]{\left(#1\right)}
\newcommand{\minimizer}{x_\textsf{min}}
\newcommand{\maximizer}{x_\textsf{max}}
\newcommand{\minvalue}{f_{\textsf{min}}}
\newcommand{\twominvalue}{f_{\textsf{2min}}}
\newcommand{\maxvalue}{f_{\textsf{max}}}
\newcommand{\minvalueprime}{f'_{\textsf{min}}}
\newcommand{\maxvalueprime}{f'_{\textsf{max}}}
\newcommand{\obj}[1]{f_{\textsf{obj}}\pa{#1}}
\newcommand{\objprime}[1]{f'_{\textsf{obj}}\pa{#1}}
\newcommand{\OPUBO}{O_{\textsf{\PUBO}}}
\newcommand{\OPUBOinverse}{O_{\textsf{\PUBO}}^{-1}}

\newcommand{\RZ}[1]{\scriptstyle{R_Z}\pa{\scriptscriptstyle{#1}}}
\newcommand{\UConstraints}{U_{\textsf{Cstr}}}
\newcommand{\UConstraintsinverse}{U_{\textsf{Cstr}}^{-1}}
\newcommand{\OPCBO}{O_{\textsf{\PCBO}}}

\newcommand{\Ud}{U_{\textsf{d}}}
\newcommand{\fig}[2]{
    \begin{figure}[H]
        \centering
        \includegraphics[width=0.5\textwidth]{\imagePath#1.png}
        \caption{#2}\label{fig:#1}
    \end{figure}
}
\newcommand{\figs}[3]{
    \begin{figure*}[htbp]
        \centering
        #2
        \caption{#3}\label{fig:#1}
    \end{figure*}
}

\newcommand{\subFig}[2]{
    \begin{subcaptionblock}{0.45\textwidth}
        \includegraphics[width=\linewidth]{images/#1.png}
        \caption{#2}\label{fig:#1}
    \end{subcaptionblock}
}

\newcommand{\Fig}[1]{Figure \ref{fig:#1}}
\newcommand{\eq}[2]{
    \begin{align}
        #2
        \label{eq:#1}
    \end{align}
}

\newcommand{\wreal}{w_{\textsf{R}}}
\newcommand{\wimag}{w_{\textsf{I}}}
\newcommand{\wperp}{w^{\perp}}
\newcommand{\ketwreal}{\ket{\wreal}}
\newcommand{\ketwimag}{\ket{\wimag}}
\newcommand{\ketwperp}{\ket{\wperp}}
\newcommand{\norm}[1]{\left\lVert#1\right\rVert}

\begin{document}
    \title{Phase-Selective Amplitude Amplification for Constrained Optimization}
    \author{Massimiliano Cutugno}
    \date{\today}
    \maketitle

    \begin{abstract}
        This work introduces a variant of Grover amplitude amplification using stabilizer and blade qubits to improve boosting robustness across objective distributions. We give geometric intuition, example simulations, and discuss limitations. Formal performance bounds and larger-scale validation are left for future research.
    \end{abstract}

    \section{Introduction}
        Combinatorial optimization problems appear across logistics, finance, AI, and network design, but solving them exactly is computationally intractable for large instances. Most such problems reduce to Constrained Polynomial Binary Optimization (\PCBO) or its quadratic form (QUBO), which are NP-hard and remain a central focus in both theory and applications.

        Quantum computing has been proposed as a path to faster solutions. Grover’s algorithm and Grover Adaptive Search (GAS) provide quadratic improvements but still inherit exponential scaling and, in the case of GAS, require bulky QFT embeddings. NISQ-friendly approaches such as QAOA and quantum annealing avoid this overhead but rely on heuristics and may fail to preserve constraints. Analog Grover variants, which encode objective values directly as phases, bypass QFT but are distribution-sensitive and sometimes amplify non-optimal solutions.
        
        This work introduces a new Grover-style variant using stabilizer and blade qubits to make phase-based amplitude amplification robust across arbitrary distributions and to ensure that the minimizer is amplified. The runtime matches GAS asymptotically but with a smaller constant. We provide geometric intuition, example simulations, and discuss limitations.

    \section{Related work}
        Grover’s algorithm provides a quadratic speedup for unstructured search problems and is often cited as a foundational quantum approach to combinatorial optimization\cite{grover1996fast,durr1996quantum}. However, its quadratic advantage is insufficient for scaling to practical industrial instances, which often require millions or billions of variables. Grover Adaptive Search (GAS) extends Grover’s method by embedding objective functions as inequality constraints and iteratively refining the search space\cite{gilliam2021grover}. While GAS has been shown to theoretically improve performance on constrained polynomial binary optimization (CPBO) problems, it inherits Grover’s exponential scaling and requires Quantum Fourier Transform (QFT) subroutines, which are resource intensive\cite{gilliam2021grover, nam2020approximate}.

        To address near-term hardware constraints, heuristic approaches such as the Quantum Approximate Optimization Algorithm (QAOA)\cite{farhi2014quantum} and quantum annealing\cite{finnila1994quantum} have been proposed. These methods are NISQ-friendly and avoid large circuit overheads, but they may not always produce optimal solutions and often violate embedded constraints.

        More recent analog-style Grover variants attempt to bypass QFT by encoding objective values directly as phases\cite{shyamsundar2023non,koch2023variational, benchasattabuse2022amplitude}. These “lightweight” alternatives avoid heuristic search and can be simpler to implement, but their effectiveness depends heavily on the distribution of objective values. In particular, they can sometimes amplify maximizers rather than minimizers.

        Our work builds directly on this line of analog Grover methods. By introducing stabilizer and blade qubits, we eliminate distributional sensitivity and ensure that the minimizer is selectively amplified.

    \section{Bender Amplitude Amplification Algorithm}
        The algorithm proposed by this paper we refer to simply as "Blender amplitude amplification algorithm" or just "Blender" for convenience. It is just an instance of amplitude amplification, which consists of constructing a unitary oracle (we refer to it as the Blender Oracle) based off the objective function and then using it as the oracle in Grover's algorithm. The Blender oracle is composed of two different parts. The first is the constraints unitary, which defines the constraints of the \PCBO{}. The other part is the \PUBO{} --Unconstrained Polynomial Binary Optimization-- oracle, which defines the value of the objective function (\PUBO{} is associated with the \PCBO{} algorithm within other research papers). Just to note, the name \PUBO{} is also described as PUBO (Polynomial Unconstrained Binary Optimization) or HOBO (Higher Order Binary Optimization), but this paper prefers to use the acronym \PUBO{}. 

        \subsection{Notation}
            Before getting into the specifics of the algorithm, let's first lay out some notation for a \PCBO{} problem. Given some \PCBO{} problem, there are \(n\) boolean variables (variables that can either be 0 or 1), with each variable notated as $x_i$ with $i\in\mathbb{N}_{0< i \leq n}$. The whole collection of variables is notated as \(x\). The weights that define the \PCBO{} problem are notated in total as $W$. $W_i$ is used for a weight applied to variable $x_i$ when true and $W_{\overline{i}}$ for a weight applied to variable $x_i$ when false. For weight of two variables, notation $W_{i,j}$ is used for the weight applied to the conjunction of $x_i$ and $x_j$ when true, and $W_{\overline{i},j}$ for the weight applied to the conjunction of $x_i$ and $x_j$ when $x_i$ is false but $x_j$ is true (vice versa with $W_{i,\overline{j}}$). The same can be said with $W_{i,j, k}$ and so on and so forth. The objective function is notated as 
            \begin{align}
                \obj{x} &= \sum_{i=1}^{n} W_i\, x_i + \sum_{i=1}^{n} W_{\overline{i}}\, (1-x_i) \nonumber \\ 
                    &\quad + \sum_{i=1}^{n} \sum_{\substack{j = 1 \\ j \neq i}}^{n} W_{i,j}\, x_i x_j \nonumber \\
                    &\quad + \sum_{i=1}^{n} \sum_{\substack{j = 1 \\ j \neq i}}^{n} W_{\overline{i},j}\, (1-x_i) x_j \nonumber \\
                    &\quad + \sum_{i=1}^{n} \sum_{\substack{j = 1 \\ j \neq i}}^{n} W_{i,\overline{j}}\, x_i (1-x_j) \nonumber \\
                    &\quad + \sum_{i=1}^{n} \sum_{\substack{j = 1 \\ j \neq i}}^{n} \sum_{\substack{k = 1 \\ k \neq i \\ k \neq j}}^{n} W_{i,j,k}\, x_i x_j x_k \nonumber \\ 
                    &\quad + \ldots
            \end{align}

            Additionally, let's notate the minimizer (the value of interest) and maximizer of the objective function as $\minimizer$ and $\maximizer$ respectfully. For any given problem, there might be duplicate minimizers and maximizers but $\minimizer$ and $\maximizer$ be used to indicate just one of them regardless of the multiplicities.  The objective function minimum and maximum values are notated as $\minvalue = \obj{\minimizer}$ and $\maxvalue = \obj{\maximizer}$ respectfully.

            The Blender algorithm \textbf{requires} the objective function to have a range of values that are bounded by $-\pi$ and $0$, with the additional requirement that the minimizer must have the objective value to be $\minvalue = -\pi$ (to some level of precision). Fortunately, if one knows the minimum and maximum value of the objective function, one can scale the weights of the \PUBO{} problem to fit the range of $-\pi$ to $0$. Of course, for most problems, one does not know the minimum or maximum values, nor a fast way of computing them. Finding the minimum or maximum values is the subject for future work. Given the minimum and maximum values of unscaled objective function $\objprime{x}$, the scaling factor is the following:

            \begin{equation}
                \textsf{scale} = \frac{\pi}{\maxvalueprime - \minvalueprime}
            \end{equation}

            Then, one just needs to scale and translate the weights so then the whole objective function would be between $-\pi$ and $0$. Given unscaled weights $W'$, the scaled weights are the following:

            \begin{align}
                W = \{ \textsf{scale} \cdot (w' -\minvalueprime) - \pi \mid  w'\in W' \}
            \end{align}

            Without loss of generality, this paper is focusing on finding $\minimizer$ of the objective function $\obj{x}$; the same process can be applied to finding the $\maximizer$ as well. 

        \subsection{Create Constraints Unitary}
            The constraint's unitary $\UConstraints$ is fairly straightforward to implement. The unitary takes in $n$ qubits $x$ representing each of the boolean variables in the \PCBO{} problem. It also takes in a qubit called the ''target" that starts off as $\ket{0}$, but $\UConstraints$ turns it to $\ket{1}$ if the input $x$ adheres to the constraints. It also can take in a certain amount of ancilla qubits that represents intermediate boolean expressions to be used to create the value of the constraint qubit (C). The process of taking a boolean expression and translating it to $\UConstraints$ is fairly straightforward. For example, say we have three variables $x_1$, $x_2$ and $x_3$, and we want to create a constraint that says $\textsf{C} = \pa{x_1 \land x_2} \lor \pa{x_2 \land x_3}$ must be true. First, let's notate intermediate variables representing $a_1 = \pa{x_1 \land x_2}$ and $a_2 = \pa{x_2 \land x_3}$ respectfully. The unitary for this constraint is the following:

            \[
            \begin{quantikz}
                \lstick{$x_1$} & \ctrl{4}\gategroup[6, steps=5, style={dashed,rounded corners,fill=blue!20, inner xsep=2pt},background, label style={yshift=0.2cm}]{$\UConstraints$} & \qw & \qw & \qw & \qw & \qw \\
                \lstick{$x_2$} & \ctrl{3} & \ctrl{4} & \qw & \qw & \qw & \qw  \\
                \lstick{$x_3$} & \qw & \ctrl{3} & \qw & \qw & \qw & \qw \\[.40cm]
                \lstick{$\textsf{C}$} & \qw & \qw & \targ{} & \targ{} & \targ{} & \qw \\[.40cm]
                \lstick{$a_1$} & \targ{} & \qw & \ctrl{-1} & \qw & \ctrl{-1} & \qw \\
                \lstick{$a_2$} & \qw & \targ{} & \qw & \ctrl{-2} & \ctrl{-2} & \qw 
            \end{quantikz}
            \]
            
            The algorithm also requires the inverse of the constraints unitary $\UConstraintsinverse$. In the case of the example above, the inverse is the same as the constraints unitary, but with gates applied in the reverse order:

            \[
            \begin{quantikz}
                \lstick{$x_1$} & \qw \gategroup[6, steps=5, style={dashed,rounded corners,fill=blue!20, inner xsep=2pt},background, label style={yshift=0.2cm}]{$\UConstraintsinverse$} & \qw & \qw & \qw & \ctrl{4} & \qw \\
                \lstick{$x_2$} & \qw & \qw & \qw & \ctrl{4} & \ctrl{3} & \qw  \\
                \lstick{$x_3$} & \qw & \qw & \qw & \ctrl{3} & \qw & \qw \\[.40cm]
                \lstick{$\textsf{C}$} & \targ{} & \targ{} & \targ{} & \qw & \qw & \qw \\[.40cm]
                \lstick{$a_1$} & \ctrl{-1} & \qw & \ctrl{-1} & \qw & \targ{} & \qw \\
                \lstick{$a_2$} & \ctrl{-2} & \ctrl{-2} & \qw & \targ{} & \qw & \qw 
            \end{quantikz}
            \]

            Although --here-- a simple boolean expression constraint is provided, more complicated weighted equality or inequality constraints can be also created here such as $3\, x_1 \cdot x_2 + 2\, x_2 \cdot x_3 \leq 5$. The GAS paper\cite{gilliam2021grover} describes how to create these kinds of constraints optimally using QFT called ''constraint-augmented quantum dictionary"; fortunately, the process is exactly the same for the Blender algorithm and won't be discussed in this paper.

        \subsection{Create \PUBO{} Oracle}
            Given the scaled weights ($W$) of the \PUBO{} problem, one can create the \PUBO{} oracle unitary in a straight forward manner (notated as $\OPUBO$). The unitary representing the oracle requires qubit input for each variable in the \PUBO{} problem. A qubit named constraint ($\textsf{C}$) with value 1 is used to apply phase to each state, although this can be done without use of the constraint qubit if no constraints are used. An example of a \PUBO{} oracle unitary for a 3-variable \PUBO{} problem with weights $W_1$, $W_{1,\overline{2}}$ and $W_{\overline{1},2,\overline{3}}$ is the following:
            
            \[
            \begin{quantikz}
                \lstick{$x_1$} & \ctrl{3}\gategroup[4, steps=3, style={dashed,rounded corners,fill=blue!20, inner xsep=2pt},background, label style={yshift=0.2cm}]{$\OPUBO$} & \ctrl{3} & \octrl{3} & \qw \\
                \lstick{$x_2$} & \qw      & \octrl{2}  & \ctrl{2} & \qw  \\
                \lstick{$x_3$} & \qw      & \qw  & \octrl{1} & \qw \\[.5cm]
                \lstick{$\textsf{C}$} & \gate{\RZ{W_1}} & \gate{\RZ{W_{1,\overline{2}}}}  & \gate{\RZ{W_{\overline{1},2,\overline{3}}}} & \qw
            \end{quantikz}
            \]
        
        The algorithm requires the inverse of the \PUBO{} oracle as well. The inverse of the \PUBO{} oracle is the same as the \PUBO{} oracle, but with the weights negated (no need to reverse the order of the gates since phase gates are commutative). The inverse of the \PUBO{} oracle is notated as $\OPUBOinverse$. The inverse of the \PUBO{} oracle for the example above is the following:

            \[
            \begin{quantikz}
                \lstick{$x_1$} & \ctrl{3}\gategroup[4, steps=3, style={dashed,rounded corners,fill=blue!20, inner xsep=2pt},background, label style={yshift=0.2cm}]{$\OPUBOinverse$} & \ctrl{3} & \octrl{3} & \qw \\
                \lstick{$x_2$} & \qw      & \octrl{2}  & \ctrl{2} & \qw  \\
                \lstick{$x_3$} & \qw      & \qw  & \octrl{1} & \qw \\[.5cm]
                \lstick{$\textsf{C}$} & \gate{\RZ{-W_1}}   & \gate{\RZ{-W_{1,\overline{2}}}}  & \gate{\RZ{-W_{\overline{1},2,\overline{3}}}} & \qw
            \end{quantikz}
            \]
            
        \subsection{Create \PCBO{} Oracle}
            Given the \PUBO{} oracle and the constraint oracle $\UConstraints$ (and their inverses), one can create the Blender oracle unitary (notated as $\OPCBO$). The Blender oracle is the following:

            \[
            \begin{quantikz}
                \lstick{$\textsf{B}$} & \qwbundle{b} & \qw \gategroup[5, steps=4, style={dashed,rounded corners,fill=blue!20, inner xsep=2pt},background, label style={yshift=0.2cm}]{$\OPCBO$} & \ctrl{2} & \ctrl{2} & \qw & \qw  \\
                \lstick{$\textsf{S}$} & \qw & \qw & \octrl{1} & \ctrl{1} & \qw & \qw  \\
                \lstick{$x$} & \qwbundle{n} & \gate[3]{\UConstraints} & \gate[2]{\OPUBO} & \gate[2]{\OPUBOinverse} & \gate[3]{\UConstraintsinverse} & \qw \\
                \lstick{$\textsf{C}$} & \qw & \qw & \qw & \qw & \qw & \qw  \\
                \lstick{$\textsf{A}$} & \qwbundle{k} & \qw & \qw & \qw & \qw & \qw 
            \end{quantikz}
            \]

            The $b$ amount of qubits $\textsf{B}$ represents the metaphorical blade of the blender, which the higher the amount of $b$, the stronger the metaphorical blade, the better the probability of measuring out the minimizer; the intuition behind the blade qubit is given in section \ref{sec:Algorithm_Intuition}. The success probability is strongly correlated with the amount of qubits $b$, for most problems of any size $n$, $b$ can be set to 9 for about $99\%$ probability measurement of the minimum value. The trade-off with increasing $b$ is that the iteration amount of the Blender algorithm increases by a multiplicative factor of $\sqrt{2}$. One has to be careful not to scale $b$ at least asymptotically linearly in terms of $n$, as this will give no quantum advantage over classical. Fortunately (for normal-like distributions), when the space between the minimum value and the second most minimum value is not too small (more on this later), the probability of measuring the minimizer rapidly increases close to $100\%$ as one slowly increases $b$ independent of $n$; From experimental results, we pose this conjecture: the probability of measuring out $\minimizer$ is $\pa{1-\epsilon} \cdot 100\%$, where the number of blade qubits chosen is $b = O\pa{\textsf{log}\pa{\frac{1}{\epsilon}}}$. When the space between the minimum value and the second most minimum value is very small (called the band gap energy if this objective function relates to the energy levels of solid-state physics), then $b$ must be scaled logarithmically with the resolution magnitude between these two values (or linearly with the order of magnitude of the resolution). There is a metaphorical way to remember when to increase $b$: adding more blades (adding increasing $b$) for mixing objective function allows blending the objective function with values of smaller granularity, otherwise it won't blend well.

            The quantity $\textsf{S}$ is the stabilizer qubit, which is a single qubit that represents the side of the Blender where the weight phases are being loaded into ($s=\ket{0}$ represents from $-\pi$ to $0$ and $s=\ket{1}$ represents $0$ to $-\pi$); the algorithmic purpose of $\textsf{S}$ is described in section \ref{sec:Algorithm_Intuition}. The $x$, $\textsf{C}$, and $\textsf{A}$ qubits have been discussed in the previous sections.

        \subsection{Blender using Grover}
            The Blender algorithm is just Grover's algorithm using the Blender Oracle $\OPCBO$ and the Grover Diffusion Operator $\Ud$ \cite{grover1996fast} (see Appendix \ref{Appendix:Diffusion} for diffusion implementation). The algorithm is the following:
            
            \[
            \begin{quantikz}
                \lstick{$\textsf{B}$ \ket{0}} & \qwbundle{b} & \gate{H} & \gate[5]{\OPCBO}\gategroup[5, steps=2, style={dashed,rounded corners, inner xsep=2pt},background, label style={yshift=0.2cm}]{$t\,\, \textsf{iterations} \rightarrow$} & \gate[3]{\Ud} & \ \ldots\ & \qw  \\
                \lstick{$\textsf{S}$ \ket{0}} & \qw & \gate{H} & \qw & \qw & \ \ldots\ & \qw \\
                \lstick{$x$ \ket{0}} & \qwbundle{n} & \gate{H} & \qw & \qw & \ \ldots\ & \metercw[label style={inner sep=1pt}]{\minimizer} \\
                \lstick{$\textsf{C}$ \ket{0}} & \qw & \qw & \qw & \qw & \ \ldots\ & \qw \\
                \lstick{$\textsf{A}$ \ket{0}} & \qwbundle{k} & \qw & \qw & \qw & \ \ldots\ & \qw 
            \end{quantikz}
            \]

            The number of iterations $t$ is approximately $\frac{\pi}{4}\sqrt{\frac{N}{M}}$ for a sufficient high value of $b$ (usually $b=9$ is high enough), where $N=2^{n+b+1}$ is the number of possible states and $M$ is the number of minimizers (since there can be more than one minimizer that is associated with the lowest minimum value while abiding by the constraints of the problem). The value $b$ should scale as the following:

            \begin{equation}
                b = \mathcal{O}\pa{\ln{\pa{\frac{\twominvalue - \minvalue}{\maxvalue - \minvalue}}}}
            \end{equation}

            Where $\twominvalue$ is the second most minimum value of the objective function that abides by the constraints and $\twominvalue \ne \minvalue$.

    \section{Algorithm Intuition}\label{sec:Algorithm_Intuition}
        The reason why the Blender algorithm works is very similar to the reasoning why Grover's algorithm works. The difference between pure Grover and the Blender Algorithm is that state's amplitudes are not just possibly negated via the oracle (leaving all state amplitudes on the real line), but they are given a multiplicative phase that swings them into the complex plane. The boost in the algorithm is the same concept as Grovers, where diffusion operation $\Ud$ just simply reflects all amplitudes over the current average of those amplitudes (but now in the complex plane). This reflection operation gives the best amplify/boost potential to a target state's amplitude when the origin is always exactly in-line somewhere between the target state's amplitude and the mean (see \cite{koch2023variational}). Other states that do not always have this property --as Grover iterations are applied-- eventually rotate themselves into the position where the mean is close-to/exactly inline somewhere between the origin and that state's amplitude, causing that amplitude to attenuate/unboost. Therefore Grover's Algorithm --now applied on state amplitudes living in the complex plane-- creates a vortex-like force which sucks bad states into the origin making them have near-zero amplitudes, while target state amplitudes get whipped out almost if a centrifugal force was bestowed upon them, giving the algorithm a Blender-like effect. In summary, the mechanism for the metaphorical vortex-like force and centrifugal force is controlled by the mean of the amplitudes and powered by the Grover diffusion operator $\Ud$.

        To illustrate an example of this, let's first look see how we might view these amplitudes on the complex plane. Let's say we have a 3-variable \PUBO{} problem with 8 possible states. The amplitudes in a superposition state can be viewed like \Fig{superposition_state} on the complex plane.
        
        \figs{dots}{
            \subFig{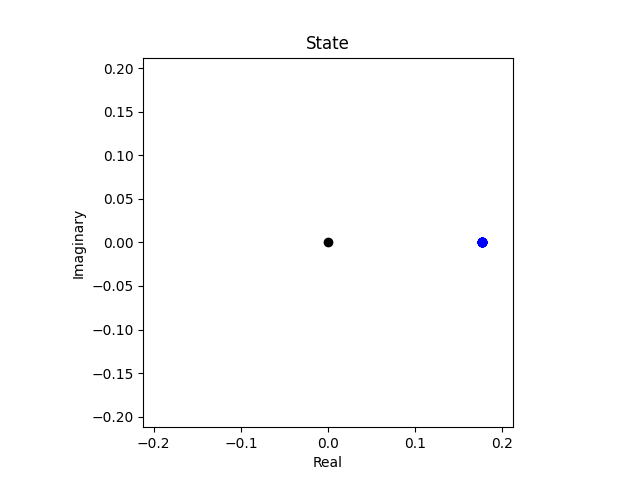}{
                This is the equal superposition state with 3-variables and 8 possible states. The blue dot represents the amplitudes of the \ket{0} through \ket{7} all sitting on-top of each other.
            }
            \subFig{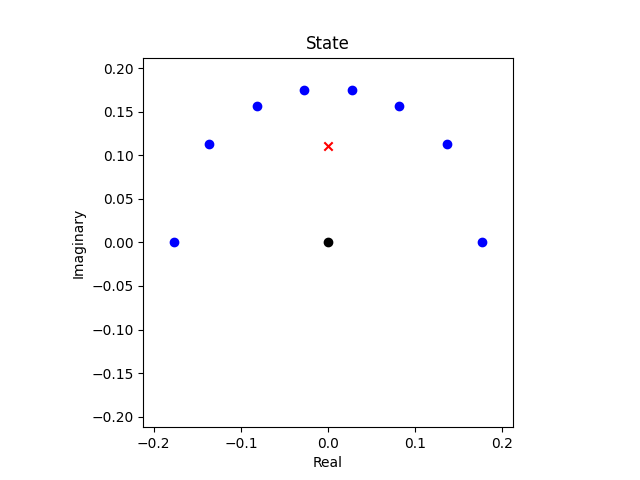}{
                The amplitudes of \Fig{superposition_state} after one application of the \PUBO{} oracle. The $\minimizer$ can be seen at phase $-\pi$ as a blue dot while the $\maximizer$ can be seen at phase $0$ as a blue dot. The other states are blue dots phased between these values and correspond to the objective function. The red x represents the mean. 
            }
            \subFig{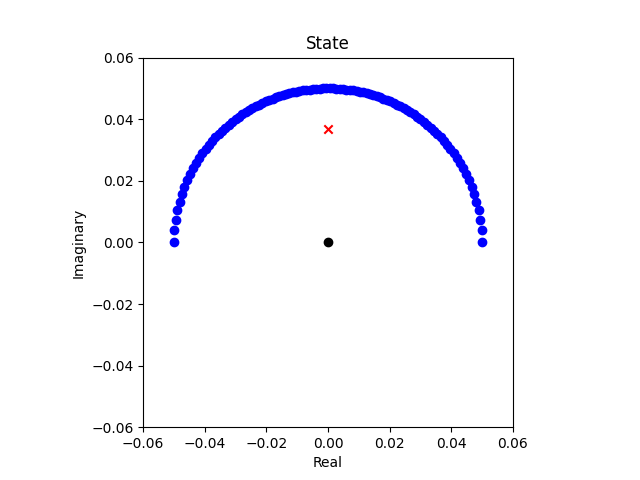}{
                The amplitudes of a superposition state of an n-variable \PUBO{} problem with more than 50 possible states after one application of the \PUBO{} oracle. The $\minimizer$ can still be seen at phase $-\pi$ as a blue dot while the $\maximizer$ can be seen at phase $0$ as a blue dot. The other states are blue dots phased between these values and correspond to the objective function.
            }
        }{
            The amplitudes of a \PUBO{} problem before and after application of \PUBO{}. The black dot is the origin, while the mean is the red x.
        }

        \figs{heatmaps}{
            \subFig{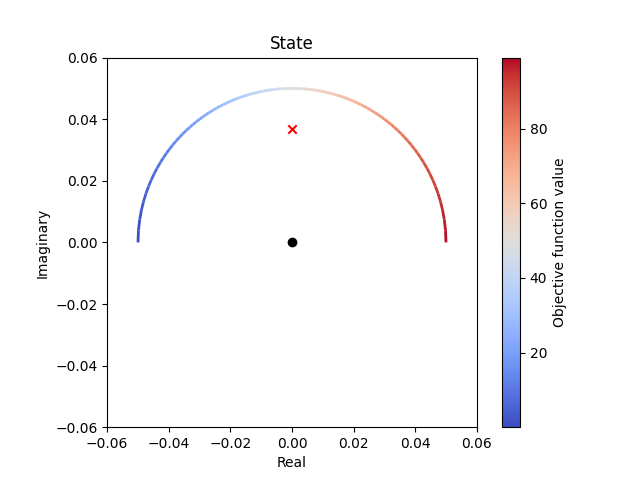}{
                This is the same superposition state as \Fig{top_dots_100} but as a heatmap.
            }
            \subFig{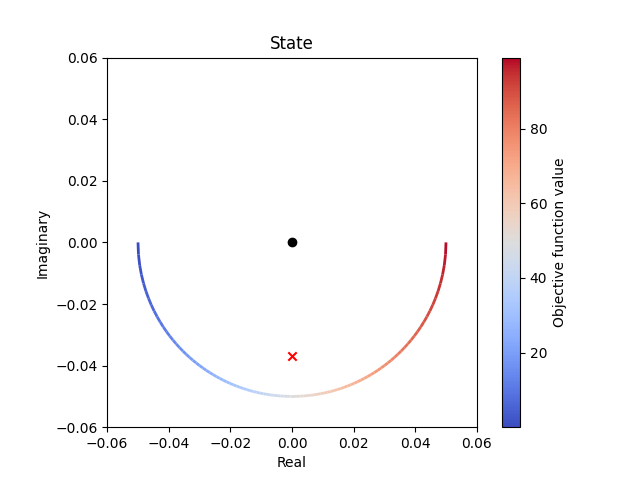} {
                This is the same superposition state as \Fig{top_heatmap_100}, but the states are phased in the negative direction; amplitudes travel clockwise.
            }

            \subFig{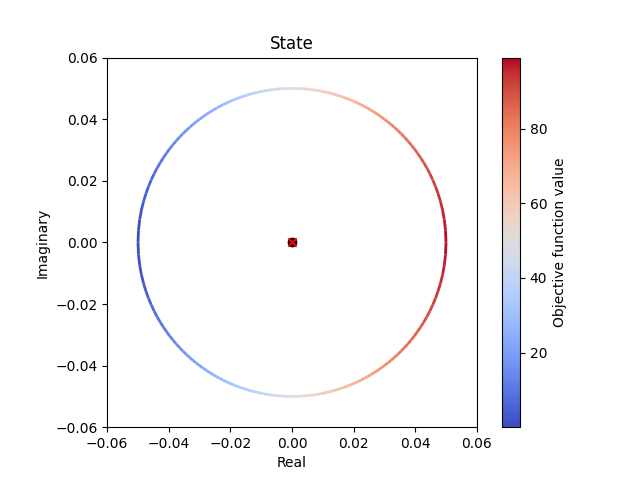} {
                This phase oracle is the combination of \Fig{top_heatmap_100} and \Fig{bot_heatmap_100}. The mean is now in-line with the $\minimizer$ and $\maximizer$ (two of each now).
            }
        }{
            Using a heatmap to show the different ways of representing the superposition state of an n-variable \PUBO{} problem with more than 50 possible states after one application of the \PUBO{} oracle. The $\minimizer$ can still be seen at phase $-\pi$ as the bluest part of the heatmap while the $\maximizer$ can be seen at phase $0$ as the reddest part of the heatmap. The other states are colors between blue and red to represent the objective function value. The red x represents the mean. The heatmap represents the objective function values of the states, with the red color representing the lowest value and the blue color representing the highest value.
        }
        \figs{blender_heatmap}{
            \subFig{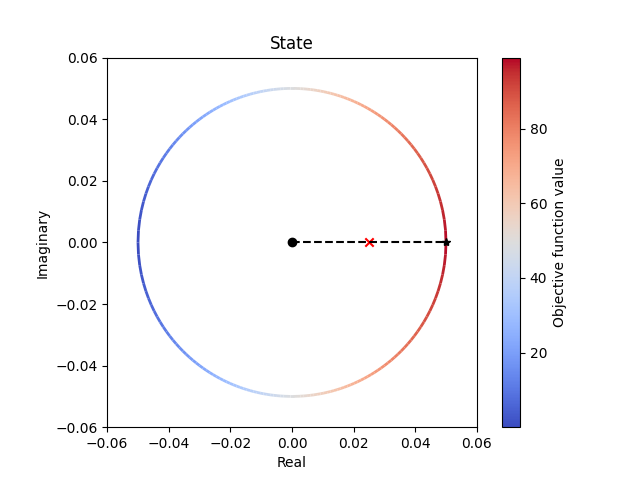}{
                The superposition state show in \Fig{both_heatmap_100} but with one blade qubit added.
            }
            \subFig{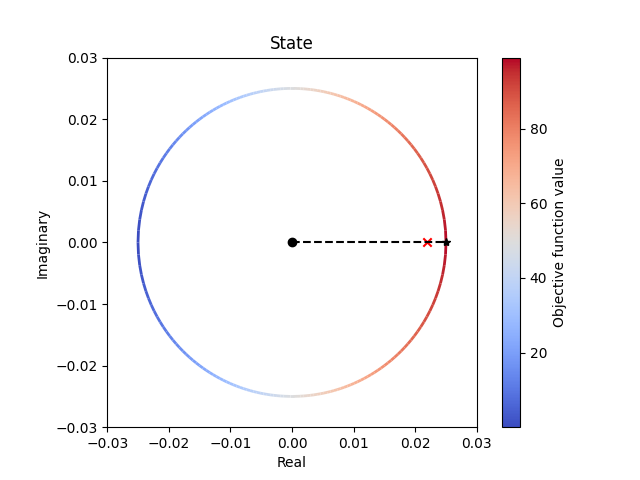}{
                The same state but with three blade qubits added.
            }
            \subFig{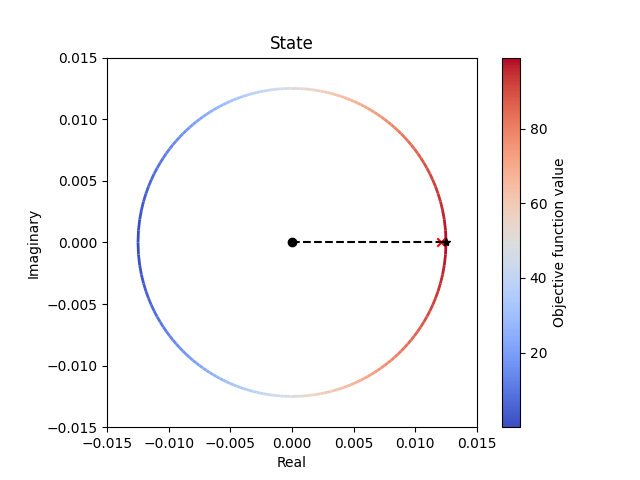}{
                The same state but with five blade qubits added.
            }
        }{
            Amplitudes after one \PUBO{} oracle with blade qubits. The blade qubit amplitudes are not phased at all and sit on the real line (phase $0$) and are shown as black stars. The blade qubit amplitudes are shown with a dashed segment to the origin only to show the metaphorical blade; the segment has no functional purpose other than the artistic representation.
        }

        Now, after application of $\OPUBO$ we see we get a phase applied to the amplitudes of the states, where the phase represents the cost of the objective function (phase $-\pi$ represents the lowest cost function value $\minvalue$ while phase $0$ represents the max $\maxvalue$). Let's say for the sake example that objective function has values distributed evenly as a perfect Gaussian, the amplitudes of the states are now on the complex plane like in \Fig{top_dots}.

        The first observation is that the mean of the amplitudes is not in-line with the $\minimizer$ and amplitude, which will provide poor amplification results. Before we go into fixing this, we would like to slightly change the plot, so we can see the objective function values of the states by applying a heatmap. If we didn't, displaying the objective function values would be hard to show when we add more points like in \Fig{top_dots_100}.

        Now, let's apply the heatmap to the complex plane to see the objective function values (ranging unweighted from 0 to 100) of the states, as if we had so many states that it is almost continuous. When the heatmap is applied to the complex plane, it looks like \Fig{top_heatmap_100}. The heatmap is a far better representation of the values of the states, so that if we add more states or apply more transformation, we can still see the objective function values.

        Fixing the problem of having the mean not in-line with the $\minimizer$ and is simple; we double the number of states and have half the states phased with the objective function with a positive imaginary value, while symmetrically across the real axis have the same copy of states phased with objective function with a negative imaginary value. That is combining \Fig{top_heatmap_100} with its mirror image \Fig{bot_heatmap_100}. The mirror image looks like \Fig{bot_heatmap_100}. The combination of the two heatmaps (which will now represent one state) looks like \Fig{both_heatmap_100}.
        
        This new state that incorporates both sides of the complex plane is easy to create. One just needs to add one mean-stabilizer qubit named $\textsf{S}$, putting it into superpostition via a Hadamard gate, then applying the $\OPUBO$ controlled with control-true by qubit $\textsf{S}$, while also applying the $\OPUBOinverse$ controlled with control-false by the $\textsf{S}$ qubit.
        
        It is now that the mean is in-line with the $\minimizer$ and $\maximizer$, but we still cannot amplify the $\minimizer$ with high probability. In this example, the mean happens to be right on top of the origin because it's a perfect symmetric Guassian; the $\minimizer$ won't boost at all. But realistically in any problem, the mean could fall anywhere on the real line between the $\minimizer$ and $\maximizer$. For optimal boosting of the minimizer, the origin must be between the $\minimizer$ and the mean and the mean must be as far as possible away from the origin.
        
        Fortunately, there is a nice remedy for the inconvenient mean placement: adding $b$ ``blade" qubits $\textsf{B}$ to the state that aren't phased at all when applying $\OPUBO$ and $\OPUBOinverse$; these states sit at phase $0$ (the real line) and are used to push the mean away from the origin. With just adding one blade qubit (shown in \Fig{blender_heatmap_100_1b}), the mean is pushed away from the origin by nice amount but still enough to get good amplification results, shown in \Fig{blender_heatmap_100_1b}.

        The more blade qubits that are added, the further the mean is pushed away from the origin as can be seen when we add 3 blade qubits or 5 blade qubits in \Fig{blender_heatmap_100_3b} and \Fig{blender_heatmap_100_5b} respectfully.

        Now, with sufficient amount of blade qubits, we use Grover's algorithm to amplify the $\minimizer$ to a high probability. After one iteration, the state looks like \Fig{blender_heatmap_100_5b_1t}.

        \clearpage
        \newpage

        \fig{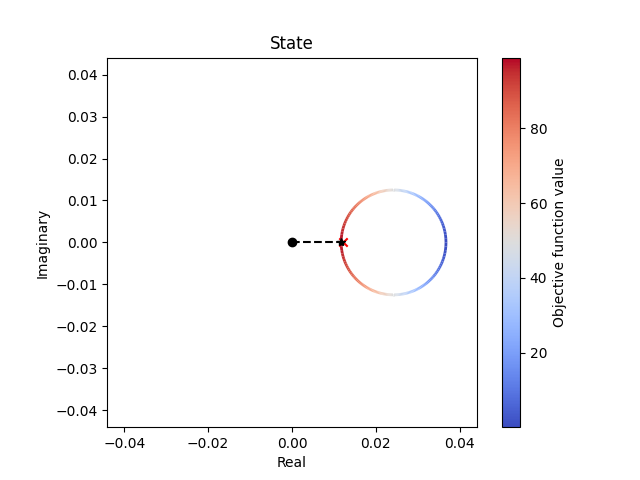}{
            The amplitudes of a superposition state of an n-variable \PUBO{} problem with more than 50 possible states after one iteration.
        }

        After 2, 11, 25, and 31 iterations, the state looks like \Fig{blender_heatmap_100_5b_2t}, \Fig{blender_heatmap_100_5b_11t}, \Fig{blender_heatmap_100_5b_25t}, and \Fig{blender_heatmap_100_5b_31t} respectfully.
        
        \figs{blender_iterations}{
            \subFig{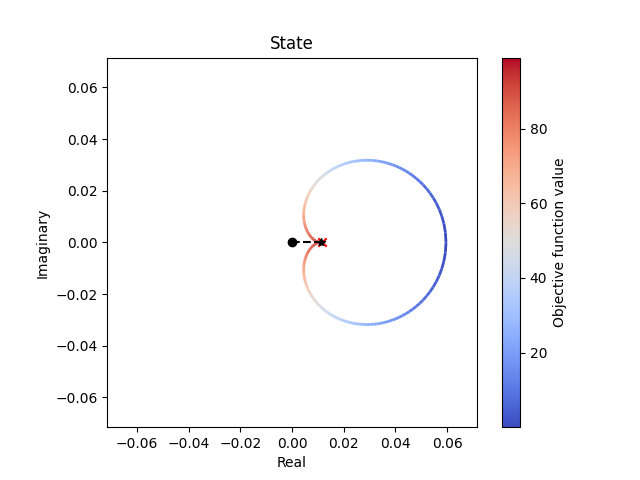}{
                $t = 2$
            }
            \subFig{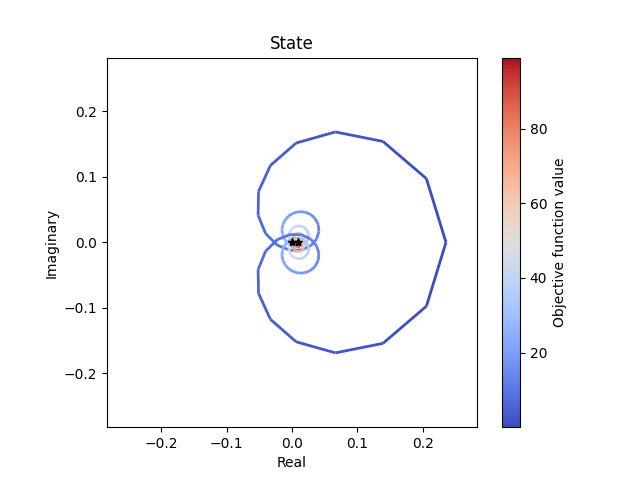}{
                $t = 11$
            }
            \subFig{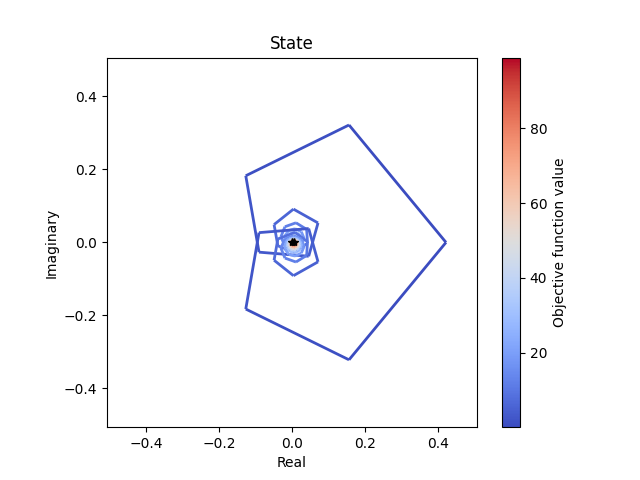}{
                $t = 25$
            }
            \subFig{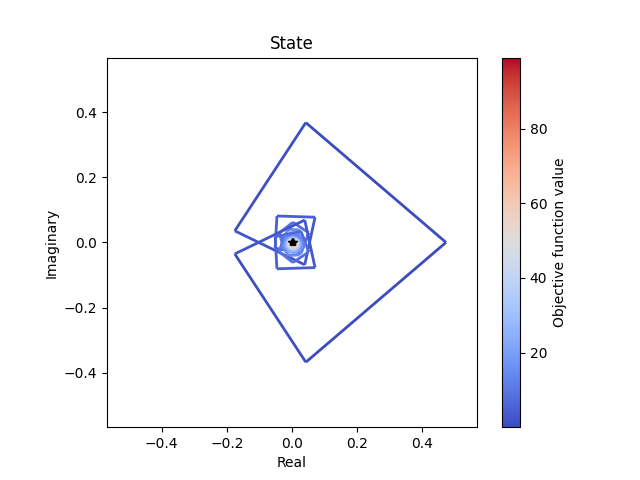}{
                $t = 31$
            }
        }{The amplitudes of a superposition state of an n-variable \PUBO{} problem with more than 50 possible states after $t$ iterations}
        
        For $n=7$ with $b=5$ the optimal probability after an optimal $t$ iterations on a perfect Gaussian is shown in \Fig{probabilities_128_5b} and is shown in log-base 2 scale in \Fig{log2_probabilities_128_5b}; it can be seen that the amplification results in high probability (around 95\%) for the minimizer (state 0 on the x axis). 

        \figs{probabilities_128}{ 
            \subFig{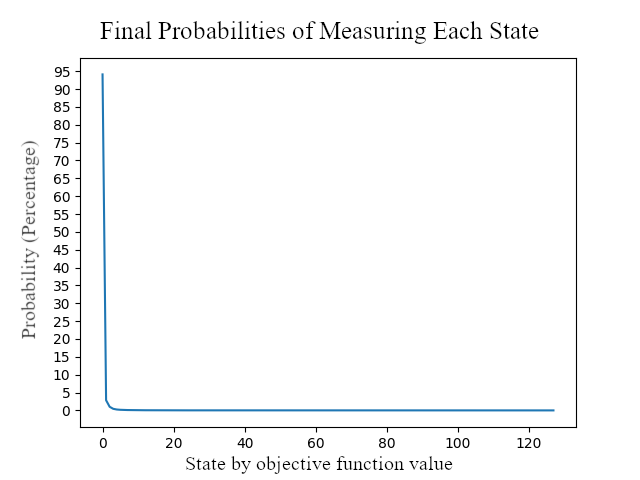}{
                Probabilities as a percentage
            }
            \subFig{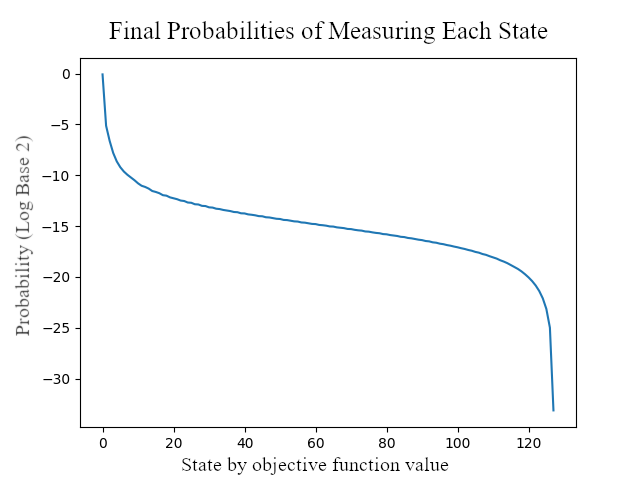}{
                Log base 2 probabilities
            }
        }{
            The probability of measuring the $\minimizer$ after the optimal amount of iterations of the Blender algorithm for a 7-variable \PUBO{} problem with 128 possible states and 5 blade qubits. The x-axis represents states ordered by their objective function values and the y-axis represents the probability of measuring out that particular state.
        }

        Everything discussed so far describes an unconstrained \PUBO{} problem. Adding constraints is a fairly easy process, just control $\OPUBO{}$ and $\OPUBOinverse$ to a qubit representing the satisfiability of constraints to apply phases only to those states that satisfy the constraints, and keep constraint-unsatisfied states with zero phase where the blade qubit states are (which make the blade metaphorically stronger). The contraints unitary that prepares this constraint qubit is what $\UConstraints$ is for. However, it is important to make sure to undo the $\UConstraints$ that was used to create the constraint qubit before applying diffusion, otherwise the diffusion will happen acros  s different subspaces rather the entire state space which has both constraint-satisfied and constraint-unsatisfied states; in other words, if constraint-satisfied and constraint-unsatisfied are in different subspaces, constraint-satisfied states can't boost off of constraint-unsatisfied states and constraint-unsatisfied states can lose its unwanted contribution of probability of measuring out. This undoing is the reason of the $\UConstraintsinverse$ unitary. Also, there is no need to differentiate gates that only specify constraints and only gates that specify phases into two operators, as one can define constraints within the phase oracle itself, as long as all the ancilla/constrain qubits are undone before applying the diffusion operator. One may even find it convenient to use multiple constraints qubits each used to add specific phases for the objective function.
    
    \clearpage
    \newpage
    
    \vspace*{20pt}
    \section{Minimum Value from QPE}
        \vspace*{-30pt}
        As discussed, the Blender algorithm requires the minimum and maximum values of the objective function to scale the weights of the \PUBO{} problem to fit the range of $-\pi$ to $0$. Just how the Quantum Phase Estimation sub-protocol was used on a vanilla Grover Unitary to pave way for the Quantum Counting algorithm\cite{brassard1998quantum}, it is natural to apply the same procedure to this modified Grover variant to possibly find $M$ and $\minvalue$ and $\maxvalue$. This section describes an unsuccessful attempt to possibly find these values using quantum phase estimation on the Blender iteration unitary when the minimum value isn't exactly on $-\pi$. In order to understand what quantum phase estimation will do on the unitary, one must understand what the measured output phases represent. To do this, we must look at the Blender algorithm (with the minimum value not at exactly $-\pi$) geometrically.

        First, we like to state a conjecture that if all states are reasonably away from $\minimizer$ (in terms of objective function value) with a reasonably high value of $b$, all other phased values have negligible significance on the algorithm. Therefore, we conjecture that the Blender algorithm's phase oracle $\OPCBO$ is approximately equivalent to just phasing the $\minimizer$ and nothing else. It's $\minimizer$ in plural because --recalling from the previous sections-- two copies for each minimizer is created: one that swings with a positive phase and one with a negative phase directed towards $\pi$/$-\pi$ (used for stabilizing the mean).
        
        Given this conjecture and model $\OPCBO$, we can see --similar to Grover's algorithm-- the Blender algorithm can be described as geometrical rotation, not in two dimensions, but in three when we consider the minimum value may not be exactly at $-\pi$. Instead of the space spanned by target state $\ket{w}$ and orthogonal state $\ket{w^{\perp}}$ (like in Grover's geometric view), the Blender algorithm can be viewed by the space spanned by the state $\ketwreal=\frac{1}{\sqrt{2}}\pa{\ket{\minimizer} + \ket{\minimizer^*}}$, $\ketwimag = \frac{1}{i\sqrt{2}}\pa{\ket{\minimizer} - \ket{\minimizer^*}}$, and $\ketwperp$, where $\ket{\minimizer}$ is the state $\minimizer$ with a positive swinging amplitude and $\ket{\minimizer^*}$ is its mirror image on the other side of the real line. $\ketwperp$ is the state that is orthogonal to the space spanned by $\ketwreal$, and $\ketwimag$.

        The the original superposition state can be spanned between the space spanned by $\ketwreal$, $\ketwimag$ and $\ketwperp$:
        \begin{figure}[H]
            \begin{tikzpicture}
                \begin{axis}[
                    view={120}{30}, 
                    grid=both,
                    axis lines=center,
                    xlabel={$\ketwimag$},
                    ylabel={$\ketwperp$},
                    zlabel={$\ketwreal$},
                    xmin=-1, xmax=1,
                    ymin=-0.5, ymax=1,
                    zmin=-1, zmax=1,
                    xlabel style={
                        anchor=east                 
                    },
                    ylabel style={
                        anchor=west                  
                    },
                ]
                    \addplot3[
                        ->, 
                        thick,
                        color=blue,
                    ] coordinates {(0,0,0) (0,{cos(20)},{sin(20)})} node[above] {$\ket{\psi}$};
                \end{axis}
            \end{tikzpicture}  
            \caption{The original superposition state $\ket{\psi}$ within the space spanned by $\ketwreal$, $\ketwimag$ and $\ketwperp$.}\label{fig:superposition_state_3d}  
        \end{figure}

        Originally, the superposition state can also exclusively be between the space spanned by $\ketwreal$ and $\ketwperp$. This can be seen from a 2D perspective in \Fig{superposition_state_3d_2}.

        \begin{figure}[H]
            \begin{tikzpicture}
                \begin{axis}[
                    view={90}{0}, 
                    grid=both,
                    axis lines=center,
                    xlabel={$\ketwimag$},
                    ylabel={$\ketwperp$},
                    zlabel={$\ketwreal$},
                    xmin=-1, xmax=1,
                    ymin=-1, ymax=1,
                    zmin=-1, zmax=1,
                    xlabel style={
                        anchor=east                 
                    },
                    ylabel style={
                        anchor=west                  
                    },
                ]
                    \addplot3[
                        ->, 
                        thick,
                        color=blue,
                    ] coordinates {(0,0,0) (0,{cos(20)},{sin(20)})} node[above] {$\ket{\psi}$};
                    \draw[->] (0, .6, 0) 
                        arc[start angle=0, end angle=20, radius=1.5cm] node[below right]{$\,\theta_s$};
                \end{axis}
            \end{tikzpicture}  
            \caption{From a 2d perspective, the original superposition state $\ket{\psi}$ within the space spanned by $\ketwreal$, $\ketwimag$ and $\ketwperp$.}\label{fig:superposition_state_3d_2}  
        \end{figure}

        After applying the Blender oracle, the superposition state then rotates around the $\ketwperp$ axis. This rotation amount is not a full 180 degrees like Grover's, but a smaller amount $\minvalue$ that is dependent on the minimum value of the objective function. This rotation is best seen from a 2D perspective now between the space spanned by $\ketwreal$ and $\ketwimag$:

        \begin{figure}[H]
            \begin{tikzpicture}
                \begin{axis}[
                    view={180}{0}, 
                    grid=both,
                    axis lines=center,
                    xlabel={$\ketwimag$},
                    ylabel={$\ketwperp$},
                    zlabel={$\ketwreal$},
                    xmin=-1, xmax=1,
                    ymin=-0.5, ymax=1,
                    zmin=-1, zmax=1,
                    xlabel style={
                        anchor=east                 
                    },
                    ylabel style={
                        anchor=west                  
                    },
                ]
                    \addplot3[
                        ->, 
                        thick,
                        color=gray,
                    ] coordinates {(0,0,0) (0,{cos(20)},{sin(20)})} node[above right] {$\ket{\psi}$};
                    \addplot3[
                        ->, 
                        thick,
                        color=blue,
                    ] coordinates {(0,0,0) ({sin(170)},{cos(20)},{cos(170)*sin(20)})} node[left] {$\OPCBO\ket{\psi}$};

                    \path (.50, 0, .25) node {$\theta=\minvalue$};

                    \draw[->, dashed] (0, {cos(20)},{sin(20)}) 
                        arc[start angle=90, end angle=220, radius=1cm];
                \end{axis}
            \end{tikzpicture}  
            \caption{The state $\ket{\psi}$ after application of the oracle $\OPCBO$}\label{fig:after_phase_3d}  
        \end{figure}

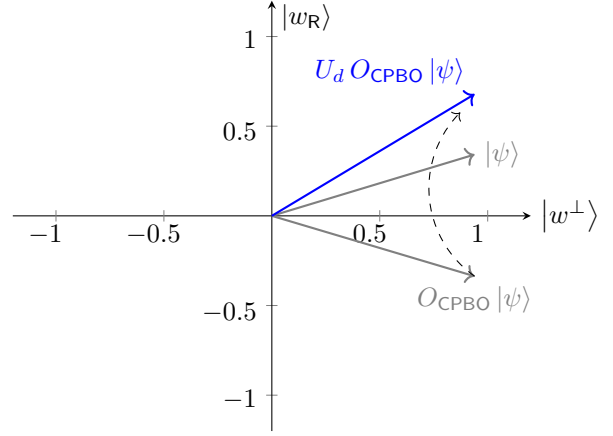
\begin{figure}[H]
            \begin{tikzpicture}
                \begin{axis}[
                    view={90}{0}, 
                    grid=both,
                    axis lines=center,
                    xlabel={$\ketwimag$},
                    ylabel={$\ketwperp$},
                    zlabel={$\ketwreal$},
                    xmin=-1, xmax=1,
                    ymin=-1.2, ymax=1.2,
                    zmin=-1.2, zmax=1.2,
                    xlabel style={
                        anchor=east                 
                    },
                    ylabel style={
                        anchor=west                  
                    },
                ]
                    \addplot3[
                        ->, 
                        thick,
                        color=gray,
                    ] coordinates {(0,0,0) (0,{cos(20)},{sin(20)})} node[right] {$\ket{\psi}$};
                    \addplot3[
                        ->, 
                        thick,
                        color=gray,
                    ] coordinates {(0,0,0) ({sin(170)},{cos(20)},{cos(170)*sin(20)})} node[below] {$\OPCBO\ket{\psi}$};
                    \addplot3[
                        ->, 
                        thick,
                        color=blue,
                    ] coordinates {(0,0,0) ({-sin(170)},{cos(20)},{sin(20) + -cos(170)*sin(20)})} node[above left] {$U_d\,\OPCBO\ket{\psi}$};

                    \draw[->, dashed] ({sin(170)},{cos(20)},{cos(170)*sin(20)}) 
                        arc[start angle=235, end angle=135, radius=1.42cm];
                \end{axis}
            \end{tikzpicture}  
            \caption{The state $\ket{\psi}$ after application of the oracle $\OPCBO$ and the diffusion operator $U_{d}$}\label{fig:after_diffusion}
        \end{figure}

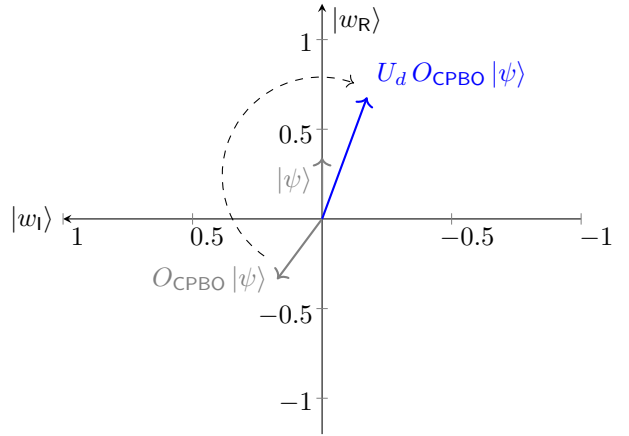
\begin{figure}[H]
            \begin{tikzpicture}
                \begin{axis}[
                    view={180}{0}, 
                    grid=both,
                    axis lines=center,
                    xlabel={$\ketwimag$},
                    ylabel={$\ketwperp$},
                    zlabel={$\ketwreal$},
                    xmin=-1, xmax=1,
                    ymin=-1.2, ymax=1.2,
                    zmin=-1.2, zmax=1.2,
                    xlabel style={
                        anchor=east                 
                    },
                    ylabel style={
                        anchor=west                  
                    },
                ]
                    \addplot3[
                        ->, 
                        thick,
                        color=gray,
                    ] coordinates {(0,0,0) (0,{cos(20)},{sin(20)})} node[below left] {$\ket{\psi}$};
                    \addplot3[
                        ->, 
                        thick,
                        color=gray,
                    ] coordinates {(0,0,0) ({sin(170)},{cos(20)},{cos(170)*sin(20)})} node[left] {$\OPCBO\ket{\psi}$};
                    \addplot3[
                        ->, 
                        thick,
                        color=blue,
                    ] coordinates {(0,0,0) ({-sin(170)},{cos(20)},{sin(20) + -cos(170)*sin(20)})} node[above right] {$U_d\,\OPCBO\ket{\psi}$};

                    \draw[->, dashed] ({sin(170) + 0.05},{cos(20)},{cos(170)*sin(20) + 0.13}) 
                        arc[start angle=235, end angle=70, radius=1.3cm];
                \end{axis}
            \end{tikzpicture}  
            \caption{The state $\ket{\psi}$ after application of the oracle $\OPCBO$ and the diffusion operator $U_{d}$ from another angle}\label{fig:after_diffusion_2} 
        \end{figure}

        The final rotation that marks a single iteration of the Blender algorithm is the application of the Grover diffusion operator $\Ud$. This rotation is hard to see in just 2D, so two different 2D perspectives are shown in \Fig{after_diffusion} and \Fig{after_diffusion_2}.

        Just like Grover's algorithm, the rotation angle of the whole iteration itself can be obtained, since a transformation of multiple rotations can be simplified into one single rotation. Because this rotation happens to be in 3D, the axis of rotation is a value of interest as well. Appendix \ref{Appendix:Rotation} shows how the axis and angle of rotation can be calculated using quaternions for one iteration of the Blender algorithm. The axis of rotation is:

        \eq{blender_axis_display}{
            \vec{r_{\textsf{axis}}} &= \begin{bmatrix}
                \hat{a}_{\wperp} \\
                \hat{a}_{\wreal} \\
                \hat{a}_{\wimag} \\    
            \end{bmatrix} \nonumber\\
            &= \frac{1}{\sin\pa{\frac{\theta_\textsf{blend}}{2}}}\begin{bmatrix}
                \sqrt{\frac{N-M}{N}}\cos\pa{\frac{\minvalue}{2}} \\
                \sqrt{\frac{M}{N}}\cos\pa{\frac{\minvalue}{2}} \\
                -\sqrt{\frac{M}{N}}\sin\pa{\frac{\minvalue}{2}} \\  
            \end{bmatrix}
        }
        as well as the angle of rotation:

        \eq{blender_angle_display}{
            \sin\pa{\frac{\theta_\textsf{blend}}{2}} = \sqrt{1-\frac{N - M}{N}\sin^2\pa{\frac{\minvalue}{2}}}
        }

        Of course, now, if one were to use phase estimation on the Blender iteration unitary, the measured phase would be the angle of rotation $\theta_\textsf{blend}$, which can be used to find the minimum value of the objective function $\minvalue$, as this angle is an eigenvalue of this iteration unitary (since it's just a rotation). This process is similar to how the quantum counting algorithm works.
        
        Unfortunately, the axis of rotation $\vec{r_{\textsf{axis}}}$ has an eigenvalue of its own of a value of $1$ within the same unitary of the Blender iteration, and the phase estimation algorithm has a large chance of measuring out this useless value instead. The reason why the axis of rotation has a large probability of measuring out is that the axis of rotation whips really close to the original superposition for even the slightest $\minvalue$ off of $-\pi$, and the phase estimation algorithm takes in a superposition of eigenvectors of the unitary. Given that $\minvalue$ is off enough from $-\pi$ and assume that $M$ is a significant smaller than $N$, the axis of rotation is close to this input superposition, therefore it has a high chance of measuring out this eigenvalue that corresponds to the eigenvector that is the axis of rotation.

        To see that the axis of rotation is close to the input superposition, we can calculate the angle between them using the dot product.

        The superposition state in this span of $\ketwreal$, $\ketwimag$ and $\ketwperp$ can be represented as the vector:

        \eq{superposition_state_vector} {
            \ket{\psi} &= \begin{bmatrix}
                \sqrt{\frac{N - M}{N}} \\
                \sqrt{\frac{M}{N}} \\
                0 \\
            \end{bmatrix}
        }
        The angle between the superposition state and the axis of rotation is then:

        \eq{dot_product_axis}{
            \cos&\pa{\theta_{\textsf{inter}}} = \frac{ \vec{r_{\textsf{axis}}}\cdot\ket{\psi} }{ \norm{\vec{r_{\textsf{axis}}}} \norm{\ket{\psi}} } \nonumber\\
            &= \frac{1}{\sin\pa{\frac{\theta_\textsf{blend}}{2}}} \begin{bmatrix}
                \sqrt{\frac{N-M}{N}}\cos\pa{\frac{\minvalue}{2}} \\
                \sqrt{\frac{M}{N}}\cos\pa{\frac{\minvalue}{2}} \\
                -\sqrt{\frac{M}{N}}\sin\pa{\frac{\minvalue}{2}} \\  
            \end{bmatrix} \cdot \begin{bmatrix}
                \sqrt{\frac{N - M}{N}} \\
                \sqrt{\frac{M}{N}} \\
                0 \\
            \end{bmatrix} \nonumber\\
            &= \frac{\cos\pa{\frac{\minvalue}{2}}}{\sin\pa{\frac{\theta_\textsf{blend}}{2}}} \nonumber \\
            &= \pa{1+\frac{M}{N}\tan^2\pa{\frac{\minvalue}{2}}}^{-\frac{1}{2}}
        }

        As said, most problems have the factor $\frac{M}{N} \ll 1$. So when $\minvalue$ is off of $-\pi$ by a small amount, the $\tan$ factor is of significance, turning the $\cos\pa{\theta_{\textsf{inter}}}$ factor close to $1$. This means that the axis of rotation is close to the original superposition state, and the phase estimation algorithm has a high chance of measuring out this eigenvalue instead of the angle of rotation $\theta_{\textsf{blend}}$. This is why the quantum counting algorithm, used in the naive form to match what quantum counting/phase-estimation, is not a good method to find the minimum value of the objective function in the Blender algorithm. However, when the minimum value is exactly at $-\pi$, the axis of rotation is orthogonal to the original superposition state, and the phase estimation algorithm will measure out the angle of rotation $\theta_{\textsf{blend}}$ with high probability, yielding information about the value of $M$ needed to known how many iterations of the algorithm should be performed.
        
        Although this method of finding the minimum value of the objective function is unsuccessful, it is still a valuable insight into the Blender algorithm and how it works geometrically. This perspective also reveals that if just the base Grover algorithm has its target state off by just a little-bit from $-\pi$, the algorithm would then boost very poorly and QPE used for quantum counting won't yield the correct value of $M$. However, in both Grover and Grover Blender Algorithms, if the target state is exactly at $-\pi$, then the algorithm will boost the target state with high probability and QPE will yield the correct value of $M$.    

    \section{Hardware Requirements}
        As can be seen from the algorithm, gate fidelity and gates that have precise phases are important (the precision can be calculated but is not specified in this paper). The Blender algorithm requires a quantum computer that has fault tolerance; the algorithm is not considered NISQ friendly and requires error-correction and long coherence times.
    
    \section{Conclusion and Future Research}
        
        It's debatable --when considering all factors-- whether the Blender algorithm has a significant advantage over GAS. For one, the Blender algorithm requires additional blade qubits adding to an increased iteration amount, increasing the runtime. This in combination with the fact that one requires knowledge of $\minvalue$ to some precision. GAS doesn't require either of these things and the iteration amount only depends on how many binary variables the problem has, but at the cost of measuring out multiple quantum computations in sequence and QFT required within it's Grover oracle to find the minimum value. Also, if $\minvalue$ is known, then GAS only needs to perform one quantum measurement (rather than multiple in the general case) with the condition $\obj{x} = \minvalue$.

        To improve the applicability of Blender-Grover, following research can answer questions like: is there a way to find or approximate $\minvalue$ through other means within quantum? Would this hypothetical method's runtime, in combination with the Blender algorithm runtime, more efficient than GAS? This is a question that should be explored in future research.

        Another interesting direction of future research is to try to modify the Blender algorithm to not only boost the minimizer state that has objective value $\minvalue$, but boost states that have an objective value around the minimum to try to see if a lower running time can be achieved; this thought is inspired by how Grover with multiple target states requires far fewer iterations than just one target state. A possible way to do this is to switch the direction of the phase oracle every couple of iterations (by use of a $\textsf{X}$ gate on the $\textsf{S}$ qubit before the iteration) to prevent states that are near the minimizer from rotating behind the mean from the origin, while keeping states that are not of interest in this area, as this is what attenuates the amplitudes of those states. This has been demonstrated while studying the algorithm, but hasn't been detailed in this paper as more research is needed to finalize this approach.

        The Blender algorithm can also be explored as an `amplify-by-phase" quantum sub-protocol; just like how Quantum-Phase-Estimation and Quantum Fourier Transform is a sub-protocol to many Quantum Algorithms. The ``amplify-by-phase" sub-protocol would be to amplify a state that has a specific phase, while attenuate all others by rotating all state amplitudes such that the phase of the state of interest is aligned opposite of the blade qubits' amplitudes.

        The Blender algorithm is an improved combinatorial optimization algorithm and there is still much to be explored about it. Although the Blender algorithm, similar to GAS, is a not a promising new algorithm in terms of solving scaled-up real-world optimization problems in a practical amount of time, it can practically solve instances of smaller sizes much faster than classically known techniques.

    \clearpage
    \newpage

    \section*{Appendix}
    
    \appendix
        \section{Axis and angle of rotation of one iteration}\label{Appendix:Rotation}
            The Blender algorithm can be viewed as a rotation in 3D space. The axis of rotation and the angle of rotation can be calculated using quaternions. The quaternion representation of rotating 180 degrees around the superposition state $q_s$ is:

            \eq{quat_1}{
                \hat{a}_{\wperp} &= \cos\pa{\theta_s} \nonumber \\
                \hat{a}_{\wreal} &= \sin\pa{\theta_s} \nonumber \\
                \hat{a}_{\wimag} &= 0 \nonumber \\
                q_s &= \cos\pa{\frac{\pi}{2}} \nonumber\\ 
                &\quad + \sin\pa{\frac{\pi}{2}}\pa{\hat{a}_{\wperp}\hat{i} + \hat{a}_{\wreal}\hat{j} + \hat{a}_{\wimag}\hat{k}} \nonumber\\
                &= \cos\pa{\theta_s}\hat{i} + \sin\pa{\theta_s}\hat{j}
            }
            where $\theta_s$ is the angle of rotation of the superposition state from $\ketwperp$. The quaternion representation of the $\OPCBO$ oracle $q_{\textsf{\PCBO}}$ is:

            \eq{quat_2}{
                \hat{a}_{\wperp} &= 1 \nonumber\\
                \hat{a}_{\wreal} &= 0 \nonumber\\
                \hat{a}_{\wimag} &= 0 \nonumber\\
                q_{\textsf{\PCBO}} &=\, \cos\pa{\frac{\minvalue}{2}} \nonumber\\
                &\quad + \sin\pa{\frac{\minvalue}{2}}\pa{\hat{a}_{\wperp}\hat{i} + \hat{a}_{\wreal}\hat{j} + \hat{a}_{\wimag}\hat{k}} \nonumber\\
                &= \cos\pa{\frac{\minvalue}{2}} + \sin\pa{\frac{\minvalue}{2}}\hat{i}
            }
            where $\minvalue$ is the minimum value of the objective function (which ranges from $-\pi$ ot $0$). The quaternion representation of the blended state $q_{\textsf{blend}}$ can be calculated by multiplying the quaternion representations of the superposition state and the $\OPCBO$ oracle:

            \eq{quat_3}{
                q_{\textsf{blend}} &= q_s \cdot q_{\textsf{\PCBO}} \nonumber\\
                &= -\cos\pa{\theta_s}\sin\pa{\frac{\minvalue}{2}} \nonumber\\ 
                &\quad + \cos\pa{\theta_s}\cos\pa{\frac{\minvalue}{2}}\hat{i} \nonumber\\
                &\quad + \sin\pa{\theta_s}\cos\pa{\frac{\minvalue}{2}}\hat{j} \nonumber\\
                &\quad - \sin\pa{\theta_s}\sin\pa{\frac{\minvalue}{2}}\hat{k}
            }
            From Grover, we know that the $\cos$ and $\sin$ of the superposition angle is:

            \eq{cos_sin_s}{
                \cos\pa{\theta_s} &= \sqrt{\frac{N - M}{N}}\\
                \sin\pa{\theta_s} &= \sqrt{\frac{M}{N}}
            }
            Given the Blender quaternion $\theta_{\textsf{blend}}$ and the trigonometric functions, the angle of rotation of the one iteration of the Blender algorithm $\theta_\textsf{blend}$ can be calculated:

            \eq{blender_angle}{
                \cos\pa{\frac{\theta_\textsf{blend}}{2}} &= -\cos\pa{\theta_s}\sin\pa{\frac{\minvalue}{2}} \nonumber\\
                &= -\sqrt{\frac{N - M}{N}}\sin\pa{\frac{\minvalue}{2}} \\
                \sin\pa{\frac{\theta_\textsf{blend}}{2}} &= \sqrt{1-\cos^2\pa{\theta_s}\sin^2\pa{\frac{\minvalue}{2}}} \nonumber\\
                &= \sqrt{1-\frac{N - M}{N}\sin^2\pa{\frac{\minvalue}{2}}}
            }
            Finally, the normalized axis of rotation of one iteration $\vec{r_{\textsf{axis}}}$ can be calculated:

            \eq{blender_axis}{
                \vec{r_{\textsf{axis}}} &= \begin{bmatrix}
                    \hat{a}_{\wperp} \\
                    \hat{a}_{\wreal} \\
                    \hat{a}_{\wimag} \\    
                \end{bmatrix} \nonumber\\
                &= \frac{1}{\sin\pa{\frac{\theta_\textsf{blend}}{2}}}\begin{bmatrix}
                    \cos\pa{\theta_s}\cos\pa{\frac{\minvalue}{2}} \\
                    \sin\pa{\theta_s}\cos\pa{\frac{\minvalue}{2}} \\
                    -\sin\pa{\theta_s}\sin\pa{\frac{\minvalue}{2}} \\    
                \end{bmatrix} \\
                &= \frac{1}{\sin\pa{\frac{\theta_\textsf{blend}}{2}}}\begin{bmatrix}
                    \sqrt{\frac{N-M}{N}}\cos\pa{\frac{\minvalue}{2}} \\
                    \sqrt{\frac{M}{N}}\cos\pa{\frac{\minvalue}{2}} \\
                    -\sqrt{\frac{M}{N}}\sin\pa{\frac{\minvalue}{2}} \\    
                \end{bmatrix}
            }

    \section{Grover Diffusion operator}\label{Appendix:Diffusion}
        The Grover diffusion operator $\Ud$ operating on $K$ qubits $\psi$ can be represented as:
        \[
        \begin{quantikz}[wire types={q,q,n,q,n,q}]
            \lstick{$\psi_1$} & \gate{H}\gategroup[7, steps=5, style={dashed,rounded corners,fill=blue!20, inner xsep=2pt},background, label style={yshift=0.2cm}]{$\Ud$} & \qw & \octrl{6} & \qw & \gate{H} & \qw \\
            \lstick{$\psi_2$} & \gate{H} & \qw & \octrl{5} & \qw & \gate{H} & \qw \\
            \vdots & \vdots & & & & \vdots & \\
            \lstick{$\psi_k$} & \gate{H} & \qw & \octrl{3} & \qw & \gate{H} & \qw \\
            \vdots & \vdots & & & & \vdots & \\
            \lstick{$\psi_{K-1}$} & \gate{H} & \qw & \octrl{1} & \qw & \gate{H} & \qw \\
            \lstick{$\psi_K$} & \gate{H} & \gate{X} & \gate{Z} & \gate{X} & \gate{H} & \qw
        \end{quantikz}
        \]

    \clearpage
    \newpage
    
    \printbibliography

\end{document}